# Strong chiral dichroism in above threshold ionization and ionization rates from locally-chiral light

Ofer Neufeld[1,2,*], Hannes Hübener[1], Angel Rubio[1,3], Umberto De Giovannini[1,4]

[1]*Max Planck Institute for the Structure and Dynamics of Matter, Hamburg, Germany, 22761.*
[2]*Physics Department and Solid State Institute, Technion - Israel Institute of Technology, Haifa, Israel, 32000.*
[3]*Center for Computational Quantum Physics (CCQ), The Flatiron Institute, New York, NY, USA, 10010.*
[4]*IKERBASQUE, Basque Foundation for Science, E-48011, Bilbao, Spain.*
*Corresponding author: E-mail: ofer.neufeld@gmail.com

We derive here a new highly selective photoelectron-based chirality-sensing technique that utilizes 'locally-chiral' laser pulses. We show that this approach results in strong chiral discrimination, where the standard forwards/backwards asymmetry of photoelectron circular dichroism (PECD) is lifted. The resulting dichroism is much larger and more robust than conventional PECD, is found in all hemispheres, and is not symmetric or antisymmetric with respect to any symmetry operator. Remarkably, a CD of up to 10% survives in the angularly-integrated above-threshold ionization (ATI) spectra, and of up to 5% in the total ionization rates. We demonstrate these results through *ab-initio* calculations in the chiral molecules Bromochlorofluoromethane, Limonene, Fenchone, and Camphor. We also explore the parameter-space of the locally-chiral field and show that the observed CD is strongly correlated to the degree of chirality of the light, validating it as a measure for chiral-interaction strengths. Our results pave the way for highly selective probing of ultrafast chirality in ATI, can potentially lead to all-optical enantio-separation, and motivate the use of locally-chiral light for enhancing ultrafast spectroscopies.

## I. INTRODUCTION

Chirality is a ubiquitous naturally occurring phenomenon that plays a major role in Physics, Chemistry, and Biology. Its analysis and characterization are crucial both from a fundamental scientific point of view (e.g. for analyzing dynamical chemical processes [1], particle physics [2], and materials topology [3,4]), and from a practical aspect (e.g. in drug design [5]). Chirality is standardly analyzed with chiroptical techniques that measure the response of the medium to optical excitations. Usually, these rely on absorption circular dichroism, which requires magnetic-dipolar and/or electric-quadrupolar interactions, hence leads to very weak responses [6]. Several breakthroughs in the last decades have advanced new methods that rely solely on electric-dipole interactions, and are accordingly much more efficient. These include perturbative second-order nonlinear effects [6–11], Coulomb explosion imaging [12,13], enantiospecific state-transfer [14], photoelectron circular dichroism (PECD) in the single-photon [15–18] and multiphoton regimes [19–23] (including bi-chromatic collinear lasers [24,25]), photoexcitation circular dichroism [26,27], and HHG using bi-chromatic non-collinear lasers [28,29].

Within this 'zoo' of methods, PECD has distinguished itself as a particularly effective technique that leads to robust enantio-sensitive chiral signals on the order of 1-15% from variable targets, and which can also be applied to probe ultrafast chirality [18,30–32]. However, this technique is technically challenging because it requires measuring the angularly-resolved photoelectron spectrum (PES). This fundamental constraint arises because chiral signals in standard PECD appear only as forwards/backwards asymmetries in the photoemission, whereas the angularly-integrated PES is independent of the medium's handedness. An alternative technique that supports chiral dichroism (CD) in angularly-integrated above-threshold ionization (ATI) [33,34], as well as in total ionization rates, would pave the way for simpler realizations for probing chirality and ultrafast dynamics, as well as for chirality control and manipulation.

Here we re-formulate PECD with non-collinear and bi-chromatic laser pulses that are 'locally-chiral' [29,35]. The electric field generated by locally-chiral light carries nonzero chirality-density within the electric-dipole approximation, and is highly effective for enantio-selectivity. We show that angularly-resolved photoelectron spectra driven by locally-chiral light leads to robust chiral signals of up to 15% that are no longer forwards/backwards asymmetric. Instead, dichroism arises in all hemispheres (i.e. forwards/backwards, up/down, left/right), and it is not purely symmetric or anti-symmetric; that is, the resulting photoemission is in itself a chiral object. This fundamental aspect leads to CD of up to 10% that survives angular-integration in ATI spectra, and up to 5% in the total ionization rates. We demonstrate these effects and their generality through *ab-initio* calculations in the chiral molecules: Bromochlorofluoromethane, Limonene, Fenchone, and Camphor. We also explore the correspondence between light's degree of chirality (DOC) [35], and the chiral-signal conversion efficiency, finding a strong correlation between the two. This result supports the use of the newly derived DOC as an effective measure to quantify chiral-light-chiral-matter interactions. The new approach is expected to be highly useful for high-precision ultrafast enantio-sensing, and could potentially lead to all-optical enantio-separation.

## II. METHOD FORMULATION

We begin by describing the optical beam configuration that leads to locally-chiral light pulses, which employs the following ω-2ω two-beam geometry:

$$\mathbf{E}(t) = E_0 A(t) \operatorname{Re}\left\{ e^{i\omega t + i\eta}\hat{e}_1 + \Delta e^{2i\omega t}\hat{e}_2 \right\} \tag{1}$$

where $E_0$ is the ω electric field amplitude, $\Delta$ is the amplitude ratio between the beams, $\eta$ is a relative phase, $\hat{e}_{1,2}$ are unit vectors along the polarization direction of each beam (each beam is elliptical with ellipticities $\varepsilon_{1,2}$ and elliptical major axis angles $\beta_{1,2}$ w.r.t the *x*-axis), $A(t)$ is a dimensionless envelope function (taken in calculation to be trapezoidal with 2-cycle rise and drop sections and a 4-cycle flat-top), and we have applied the dipole approximation neglecting spatial degrees of freedom. Eq. (1) describes two noncollinear elliptically poalrized laser beams of frequencies ω and 2ω, respectively, which are focused together into a randomly-oriented chiral medium (as illustrated in Fig. 1). Note that the field in eq. (1) reduces to standard monochromatic circularly-polarized light (CPL) upon substituting $\alpha=0$, $\Delta=0$, $\varepsilon_1=1$, which allows comparing PECD obtained from locally-chiral light to the standard CPL regime.

The interaction of the laser field in eq. (1) with chiral molecules is described here using an approach that is based on time-dependent density functional theory (TDDFT) [36], in a real-space and real-time formulation using the octopus code [37–39]. This approach is extensively described in refs. [37–39], and is a non-perturbative *ab-initio* method that in principle includes both electron-ion and electron-electron interactions. We delegate technical details about the method to the appendix. For simplicity, calculations in the main text employ the single-active electron approximation (i.e. all deeper Kohn Sham states are kept frozen during propagation), which has been proven very effective in PECD [21,24,25,40]. To calculate the angularly-resolved PES, we employ here the surface flux method t-surff [41–45]. Calculations are performed consecutively for varying molecular orientations of both enantiomers to obtain full orientation averaging (see appendix A for details).

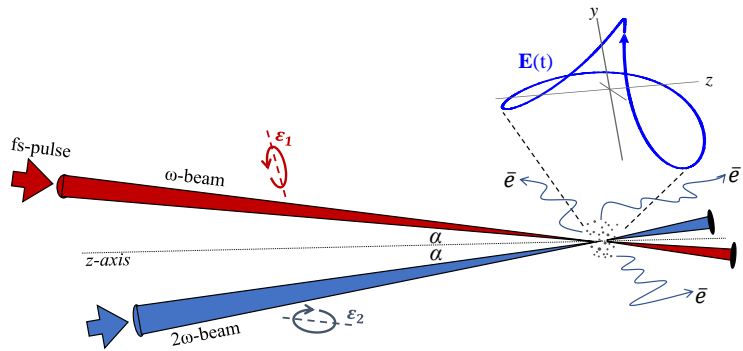

FIG. 1. Illustration of locally-chiral laser field configuration that is generated by ω-2ω biochromatic noncollinear laser pulses. Here two fs-beams of carrier frequencies ω and 2ω are focused into a gas of randomly oriented chiral molecules with an opening angle 2*α*. Each beam is elliptical with ellipticities $\varepsilon_{1,2}$, and possibly different elliptical major axes. The blue Lissajou represents the total electric field at the focus, which exhibits no improper rotational symmetries, rendering it 'locally-chiral'. The strong-field photo-ionizes electrons that are measured in an angular- and momentum-resolved fashion.

In order to put the new scheme into perspective, we first re-formulate the main physical observable of PECD, which is a CD observed in the angularly-resolved PES from mirror-image enantiomers. Theoretically, such a dichroism is obtained by subtracting the orientation-averaged PES calculated from both enantiomers that interact with the same CPL, and integrating along one axis (which experimentally occurs in velocity map imaging (VMI) [6]):

$$PECD_{CPL}(k_x, k_z) = \frac{P_R(k_x, k_z) - P_S(k_x, k_z)}{\max\{P_R(k_x, k_z)\}} \tag{2}$$

where $P_{R/S}(k_x,k_z)$ is the momentum-resolved photoelectron distribution after integration along the transverse $y$-axis from $R/S$ enantiomers, respectively, and the $z$-axis is the propagation axis of the laser pulse. Note that we have normalized the PECD to the maximal power obtained in $P_R(k_x,k_z)$, which for the CPL case is identical for both enantiomers. This naturally limits PECD to have theoretical bounds from -200 to 200%. We also note that enantiomeric exchange in eq. (2) ($R \rightarrow S$) is equivalent to exchanging the helicity of the CPL, i.e. the chiral dichroism is equivalent to the circular dichroism. Importantly, eq. (2) exhibits exact forwards/backwards asymmetry in the PECD; that is, upon exchanging the medium's handedness ($R \rightarrow S$) one finds a similar response with an opposite sign along $k_z$ [6]. The physical origin of this effect is a symmetry exhibited by any collinear laser beam – the electric field is symmetric under the transformation $E_z \rightarrow -E_z$ within the dipole approximation [46]. Major consequences of this asymmetry are: (i) no CD is observed in the $xy$ plane (i.e. in up/down or left/right hemispheres), and (ii), no CD survives angular-integration.

As opposed to the CPL case, we now formulate the main physical observables of the new approach to PECD using locally-chiral light. Here, one may define PECD with respect to any hemisphere (or Cartesian plane):

$$PECD(k_i, k_j) = 2\frac{P_R(k_i, k_j) - P_S(k_i, k_j)}{\max\{P_R(k_i, k_j)\} + \max\{P_S(k_i, k_j)\}} \tag{3}$$

where $i$ and $j$ denote cartesian indices, and eq. (3) is still bound from -200 to 200%, though now the maximal values of $P_R(k_i,k_j)$ and $P_S(k_i,k_j)$ in the denominator are not necessarily identical. This is a consequence of the light's local-chirality, which breaks the forwards/backwards asymmetry. In fact, following the structure of the light field [29,35], the resulting function $PECD(k_i,k_j)$ does not exhibit any particular symmetry relation – it contains both symmetric and anti-symmetric parts with respect to spatial reflections or enantiomeric exchange. It is also important to point out that here the chiral dichroism is no longer equivalent to circular dichroism due to the superposition structure of the field in eq. (1).

Due to the chiral nature of the PES, it is also appropriate to discuss the CD of the full photoelectron distribution:

$$PECD(\mathbf{k}) = 2\frac{P_R(\mathbf{k}) - P_S(\mathbf{k})}{\max\{P_R(\mathbf{k})\} + \max\{P_S(\mathbf{k})\}} \tag{4}$$

where $\mathbf{k}$ is the three-dimensional outgoing photoelectron momentum, and $P_R(\mathbf{k})$ contains the full PES. It is notable that for the CPL case, eq. (4) still only leads to CD due to an exchange $k_z \rightarrow -k_z$, where for locally-chiral light any possible exchange of momentum could lead to CD. While the object in eq. (4) is very difficult to experimentally resolve, it can be much more revealing towards the possible enantio-selectivity potential of PECD. Particularly, one can take the maximal value of the CD as a measure for the selectivity: $PECD_{max}=\max\{PECD(\mathbf{k})\}$. We utilize this quantity in order to compare between PECD in different conditions, and from different molecules.

Lastly, two more vital quantities should be defined. First, since locally-chiral light breaks all symmetry relations for photoemission between enantiomers, we can expect CD to survive angular-integration:

$$ATI_{CD}(\varepsilon) = 2\frac{ATI_R(\varepsilon) - ATI_S(\varepsilon)}{\max\{ATI_R(\varepsilon)\} + \max\{ATI_S(\varepsilon)\}} \tag{5}$$

where $ATI_{CD}(\varepsilon)$ is the CD obtained at the photoelectron energy $\varepsilon$, and $ATI_{R/S}(\varepsilon)$ are the individual ATI spectra from each enantiomer. For simplicity, we have normalized eq. (5) by the maximal photoelectron energy-resolved yield, which gives a good estimate to the size of the CD. Alternatively, one may normalize the ATI CD per each peak, as is done for instance in HHG [28,29,47–49], which can lead to overall larger CD values. Second, by integrating over the energy in eq. (5) we obtain the CD in total photoelectron yield:

$$I_{CD} = 2\frac{I_R - I_S}{I_R + I_S} \tag{6}$$

Here $I_{CD}$ is a scalar that is normalized from -200 to 200% that indicates the total excess electrons ionized from one enantiomer compared to the other. We emphasize that both $ATI_{CD}(\varepsilon)$ and $I_{CD}$ are strictly zero for CPL. In fact, they are strictly zero for more complex bi-chromatic fields [24,25], or for any light that does not possess local-chirality.

## III. NUMERICAL RESULTS

Having formulated the main physical observables of the new theoretical framework for PECD, we turn to practical calculations. We begin our analysis with the smallest stable chiral molecule, Bromochlorofluoromethane (CBrClFH). We calculate the PES from R- and S-CBrClFH driven by the locally-chiral field in eq. (1) (for numerical details see appendix A). The optical beam parameters are chosen according to a maximally-chiral configuration that was predicted in ref. [35] to maximize light's DOC, which is intuitively expected to yield relatively large chiral signals. Figures 2(a-c) show the resulting PECD in all cartesian planes (according to eq. (3)), which exhibit strong CD in all hemispheres and reaches a maximal value of 11.51% (this can be compared to a value of 2.13% obtained from CPL in similar conditions, see appendix B for details). A striking feature here is that Figs. 2(a-c) exhibit no symmetry relations, i.e. the standard forwards/backwards asymmetry of PECD is broken. This is a direct consequence of the use of locally-chiral light, which breaks the mirror image relation between enantiomers. It is worth mentioning that for this symmetry breaking to occur, pathways for photoemission must mix photons from both of the noncollinear beams that comprise the locally-chiral field.

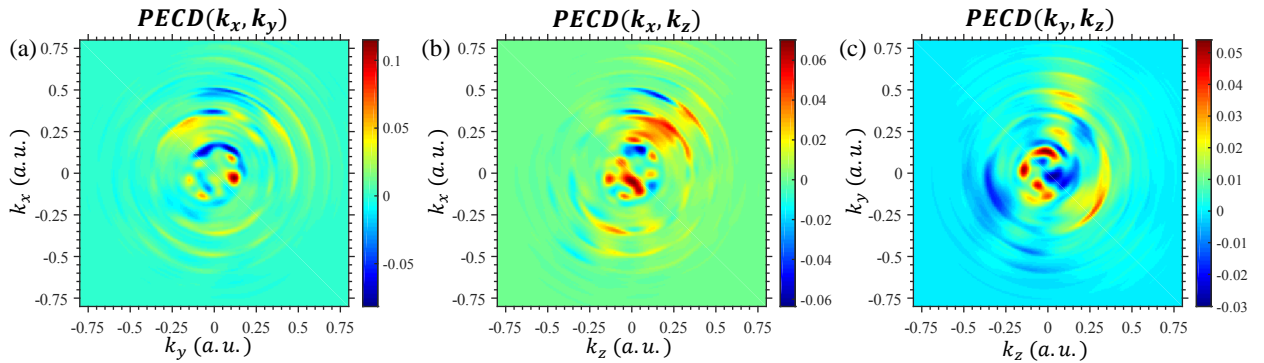

FIG. 2. PECD using locally-chiral light from CBrClFH. (a-c) $PECD(k_i,k_j)$ where $i$, $j$ are Cartesian indices. Electron momenta are given in atomic units. The locally-chiral ω-2ω laser beam set-up used is the one predicted in ref. [35] to carry maximal DOC, where the parameters in eq. (1) are set to: $\lambda$=800nm ($\omega$=1.55eV), $\varepsilon_1$=-0.37, $\varepsilon_2$=0.58, $\beta_1$=23.9º, $\beta_2$=28.7º, $\alpha$=30.1º, $\Delta$=0.77, $\eta$=0.4, and $I_0$=2×10$^{13}$ W/cm$^2$. Note that the color scales are not identical in all sub-plots.

Figure 3 presents the angularly-integrated ATI spectra from both enantiomers, and the resulting ATI CD according to eq. (5). CD of up to 4.29% is obtained for the low energy ATI peaks, and a CD of up to 2% survives up to 7 eV. In the appendix we present calculations at equivalent conditions but higher laser powers, where ATI CD of up to 7% is obtained (see Fig. 7), i.e. stronger field amplitudes generally increase the ATI dichroism, as expected (since the mixing between fields is more prominent). Note though that this can also cause the ATI CD to strongly oscillate from peak to peak (see discussion in appendix B). Overall, this broad energy-range and strong signal can be highly useful for chiral-spectroscopy.

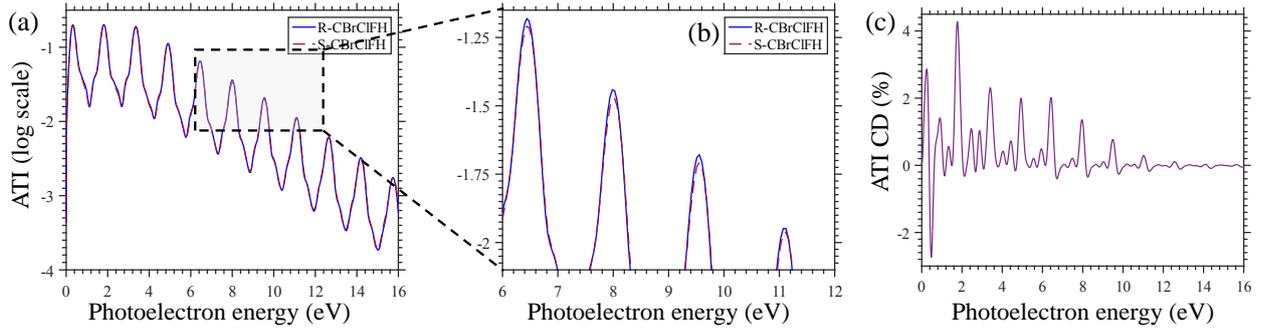

FIG. 3. ATI CD using locally-chiral light from CBrClFH. (a) ATI spectra from R- and S-CBrClFH. Inset in (b) shows magnification around 6-12 eV region, clearly showing a discrimination in the photoelectron yield around the ATI peaks. (c) Resulting ATI CD using eq. (5), normalized to the maximal ATI power. The locally-chiral $\omega$-$2\omega$ laser beam set-up is identical to that in Fig. 2.

Additional integration of the ATI spectra from both enantiomers leads to a total photoelectron yield CD (according to eq. (6)) of 2.46%. That is, when R-CBrClFH interacts with the strong field in eq. (1), electrons are photoionized at a rate that is 2.46% faster than those from S-CBrClFH. This result is somewhat unintuitive, since both molecules are mirror images of one another, and since the randomly-oriented media are fully isotropic. Still, the lack of inversion symmetry in the randomly-oriented medium, accompanied by the lack of such a symmetry in the optical set-up [46], allows one of the enantiomers to interact more efficiently with the light compared to the other enantiomer. Notably, such an approach can still be applied to time-resolved spectroscopy of dynamical processes, since the laser pulses have femtosecond durations (the FWHM in calculations is 16fs for 800nm fundamental wavelengths). It is also noteworthy that this result can pave the way for all-optical chiral-separation – it may be possible to engineer a beam configuration that fully ionizes one enantiomer, while leaving the other nearly intact; hence, after the process has concluded one of the enantiomers may be removed from the system (e.g. by Coulomb explosion [12,13]) leaving a pure sample. For this reason, it is highly important to investigate the correlation between light's physical properties like its DOC, and the chiral signal conversion efficiency, as well as to improve chiral signals by optimizing different degrees of freedom in the laser.

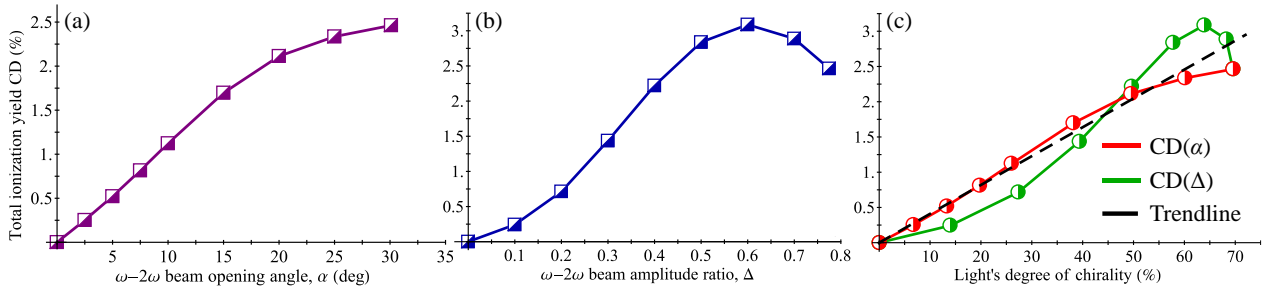

FIG. 4. Physical behavior of ionization rate CD from locally-chiral light in CBrClFH vs. laser beam parameters, and correspondence with light's DOC. (a) Total photoelectron yield CD vs. $\omega$-$2\omega$ laser beam opening angle, $\alpha$. (b) Same as (a), but as a function of the $\omega$-$2\omega$ beam amplitude ratios, $\Delta$. (c) Correspondence between total photoelectron yield CD in (a) and (b), and light's DOC, as $\alpha$ and $\Delta$ are varied (trendline shows linear regression to all data with $R^2$=0.98). In all calculations the laser beam parameters are identical to those in Fig. 2 and 3, but where only $\alpha$ or $\Delta$ are varied. DOC calculations are performed following the prescriptions in ref. [35].

With this in mind, we explore the beam parameter-space in the context of the efficiency of the chiral light-matter response, and scan the opening angle $\alpha$, and amplitude ratio $\Delta$, while calculating the total ionization rate CD. In ref. [35], the seven degrees of freedom that characterize the field in eq. (1) (e.g. ellipticities, phases, amplitudes, etc.) were optimized to yield a maximal value for its DOC, which is a theoretical measure that quantifies the extent of symmetry breaking expected by this light [29,35]. It is crucial to determine if the DOC indeed correlates to the obtained chiral signals if it is to be used for applications, which has not yet been established (i.e. as is known for the ellipticity of light in the CPL case). Results are presented in Fig. 4(a,b), where in each scan all other beam parameters are fixed to the maximal DOC configuration. Figure 4(a) shows a strong increase of the total yield CD vs. the opening angle, which is in perfect correspondence with the increase of light's DOC vs. $\alpha$ (Fig. 4(c)). Similarly, Fig. 4(b) shows a strong increase in the CD up to beam amplitude ratios of $\Delta$=0.6, where the signal maximizes at 3.09%. This increase agrees with the increase in the laser field's DOC vs. $\Delta$ (Fig. 4(c)), though here there is a slight discrepancy since light's DOC maximizes at Δ=0.77, while the chiral-response maximizes at Δ=0.6. Overall, these results support a significant correspondence between the DOC of the laser field and the chiral signal conversion efficiency (see trendline in Fig. 4(c) with $R^2$=0.98), meaning that the DOC is a useful measure for chiral-light-chiral-matter interaction strengths, and can be formally used to predict field geometries for chiral-spectroscopy or enantio-separation.

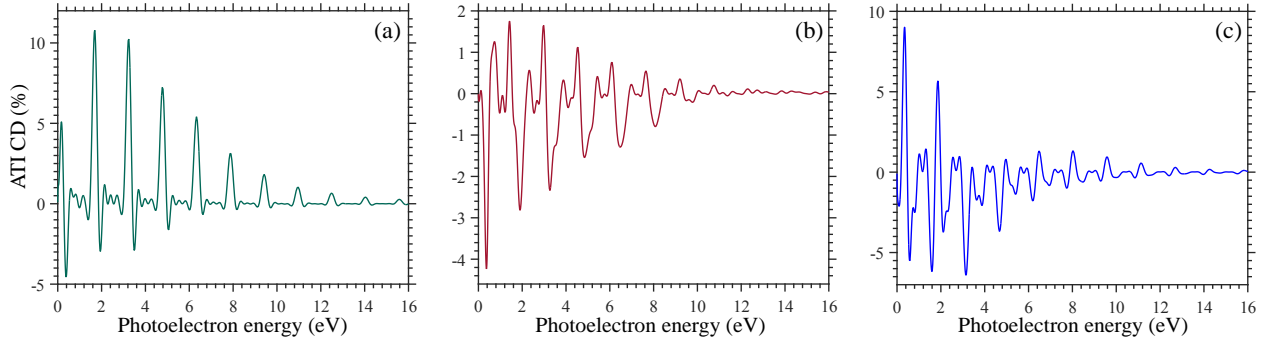

FIG. 5. ATI CD using locally-chiral light from: (a) Limonene, (b) Fenchone, (c) Camphor, respectively. Calculations are performed in similar ω-2ω optical beam settings to those in Figs. 2 and 3.

Having established the main results in CBrClFH, we demonstrate the generality of the technique by performing similar calculations in three other benchmark chiral molecules: Limonene, Fenchone, and Camphor. Figure 5(a-c) presents the corresponding ATI CD from each species in similar settings to those in Fig. 3 (see Fig. 8 for the corresponding PECD). Strong CD is observed in ATI peaks reaching as high as 10 eV from all species, where Limonene shows the strongest response with a maximal value of 10.79%. The total photoelectron yield from each species also demonstrates large CDs of up to 5%. Table 1 summarizes the different chiral observables calculated for these molecules in identical settings, as well as in the CPL case in CBrClFH.

Table 1. Summary of calculated chiral signals obtained from different chiral molecules and within different enantio-sensitive measures as described in the text. All cases utilize the same optical set-up as in Figs. 2, 3, 5, except for the CPL case, which has the same total laser power ($I_0$=4×10$^{13}$ W/cm$^2$). All values are in absolute sizes (signs have been removed).

|  | CBrClFH - CPL | CBrClFH | Limonene | Fenchone | Camphor |
|---|---|---|---|---|---|
| $\max\{PECD_{ij}\}$ | 2.13% | 11.51% | 13.61% | 12.08% | 10.89% |
| $PECD_{max}$ | 5.91% | 14.38% | 37.41% | 24.12% | 24.85% |
| $\max\{ATI_{CD}\}$ | 0 | 4.29% | 10.79% | 4.24% | 9.01% |
| $I_{CD}$ | 0 | 2.46% | 4.81% | 1.82% | 1.95% |

## IV. CONCLUSIONS AND OUTLOOK

To summarize, we have re-formulated the method of PECD to the use of noncollinear and bi-chromatic laser pulses that are locally-chiral [29,35]. We have theoretically derived the main chirality-sensitive observables for this approach, and performed *ab-initio* calculations on a set of four chiral molecules to verify its validity. The use of locally-chiral light is shown to break the forwards/backwards asymmetry of standard PECD, which leads to several new physical properties: (i) a strong CD (~5 times stronger than CPL of similar laser power) in the PES is observed in all hemispheres (i.e. in all Cartesian planes), and the photoemission is a chiral entity that does not exhibit any symmetry relation upon enantiomeric exchange. (ii) ATI CD of up to 10% is obtained in the angularly-integrated photoelectron spectra. (iii) The total ionization rate (angularly- and energy-integrated) from the orientation averaged chiral media exhibits a large CD of ~5% (note that this is on the same order of magnitude as CDs that are standardly obtained in regular PECD, see e.g. results in [17,19,20,23,25,32]). (iv) The method is independent of the femtosecond duration of the driving pulse, such that it can still be applied for exploring ultrafast dynamical processes. We have also established a strong correlation between light's DOC and the chiral signal conversion efficiency, suggesting that it can be utilized in future studies to predict ideal laser beam set-ups for chiral spectroscopy.

The exciting prospects of the new approach pave the way for using ATI and photoionization measurements for chiral discrimination. The enhanced sensitivity also means that the method is likely suitable to probe other static or dynamical properties of molecules, including valence structure and dynamical correlations. Importantly, large CDs in the total molecular ionization rates can lead the way to efficient all-optical enantio-separation *via* selective photoionization, i.e. one could potentially design a laser field that selectively photo-dissociates only one molecular handedness (DOC optimization as in ref. [35] seems like an appropriate path for future implementation). Notably, this technique can also be extended to explore chiral solids, and in particular, chiral topological effects in quantum materials. Looking forward, our work will advance ultrafast chirality spectroscopy and manipulation, and especially, motivate the use of locally-chiral light for the enhancement of existing techniques.

## ACKNOWLEDGMENTS

We thank Shaked Rozen from Weisman Institute, Israel, and Bernard Pons from Université Bordeaux, France, for helpful discussions. We acknowledge financial support from the European Research Council (ERC-2015-AdG-694097). The Flatiron Institute is a division of the Simons Foundation. O.N. gratefully acknowledges the support of the Adams Fellowship Program of the Israel Academy of Sciences and Humanities.

## APPENDIX A: NUMERICAL DETAILS

### 1. Ab-initio calculations

All DFT calculations were performed using the octopus code [37–39]. The KS equations were discretized on a Cartesian grid with spherical boundaries of radius 45 bohr, where molecular center of masses were centered at the origin. Calculations were performed using the local density approximation (LDA) with an added self-interaction correction (SIC) [50], implemented in an optimized effective potential (OEP) method (within the Krieger-Li-Iafrate (KLI) approximation [51]). This is a crucial point as adding the SIC guarantees a correct long-range KS potential that decays non-exponentially, and is required to obtain correct PECD [40]. The frozen core approximation was used for inner orbitals, which were treated with appropriate norm-conserving pseudopotentials [52]. The Kohn-Sham (KS) equations were solved to self-consistency with a tolerance $<10^{-7}$ Hartree, and the grid spacing was converged to $\Delta x=\Delta y=\Delta z=0.4$ bohr, such that the total energy per electron was converged $<10^{-3}$ Hartree. All molecular structures were relaxed $<10^{-4}$ Hartree/bohr in forces within the LDA.

For time-dependent calculations, the HOMO KS orbital was propagated with a time step $\Delta t=0.105$ a.u. (deeper levels were frozen), and by adding an imaginary absorbing potential of width 15 bohr at the boundary. The initial state was taken to be the system's ground-state. The propagator was represented by an $8^{th}$ order Taylor expansion. The grid size, absorbing potential, and time step were tested for convergence.

### 2. PECD and ATI spectra

The full PES from each molecular orientation was calculated using the t-surff method [41,42], implemented within the octopus code [43–45]. A spherical surface where flux is calculated was positioned at $r=30$ bohr, where integration was performed with a maximal angular momentum index for spherical harmonics of 40, angular grids were spanned with spacing 1°, k-grids were spanned with a spacing of $\Delta k=2\times10^{-3}$ a.u. and up to a maximal energy of 75 eV. The orientation averaged PES was calculated by trapezoidal integration as specified below, where the laser axes were repositioned and oriented with rotation matrices, and the PES was interpolated using cubic splines on the rotated grids. PECD spectra were obtained directly by subtracting the PES calculated from mirror image enantiomers. Integration over Cartesian axes and angular grids was performed using Simpson integration schemes. The total ionization rate was calculated directly from the electron density rather than integration over the ATI spectra, since this approach has improved accuracy.

### 3. Orientation averaging.

Orientation averaging was performed by spanning the angular grid with Euler angles in the *z-y-z* convention. The three Euler angles were spanned on equidistant grids with spacing $\pi/4$, leading to a total of 405 orientations and 208 irreducible orientations. Summation was performed with trapezoidal weights. The angular grid was converged against a twice reduced grid in the second Euler angle (leading to a total of 729 orientations and 464 irreducible orientations), converging the maximal ATI CD by 0.71% and the total ionization rate CD by 0.55%.

## APPENDIX B: ADDITIONAL RESULTS FROM CBrClFH

We present here additional results of calculations performed for CBrClFH that are complementary to the results presented in the main text.

First, we present PECD spectra for the CPL case (see Fig. 6), i.e. after setting $\alpha=0$, $\Delta=0$, $\varepsilon_1=1$ in eq. (1) in the main text, and after setting the laser power to $I_0=4\times10^{13}$ W/cm$^2$ to have the same total power as in calculations presented in the main text. Results show an almost perfect forwards/backwards asymmetry, as expected, with symmetric parts constituting <0.01%. The CD reaches the highest value of 2.13%, which is considerably smaller than when using locally-chiral light as shown in the main text. Additionally, PECD in the *xz* and *yz* planes are almost identical, and present almost perfect up/down left/right symmetries. Deviation from up/down and left/right symmetry is a result of the short laser pulse duration. We further verify that for the CPL case the PECD in the *xy* plane is <10$^{-3}$%, the ATI CD is <2×10$^{-3}$%, and the total ionization rates CD is <2×10$^{-4}$%. These results constitute a sanity check for the convergence of the angular grid, and for the appropriateness of the numerical approach.

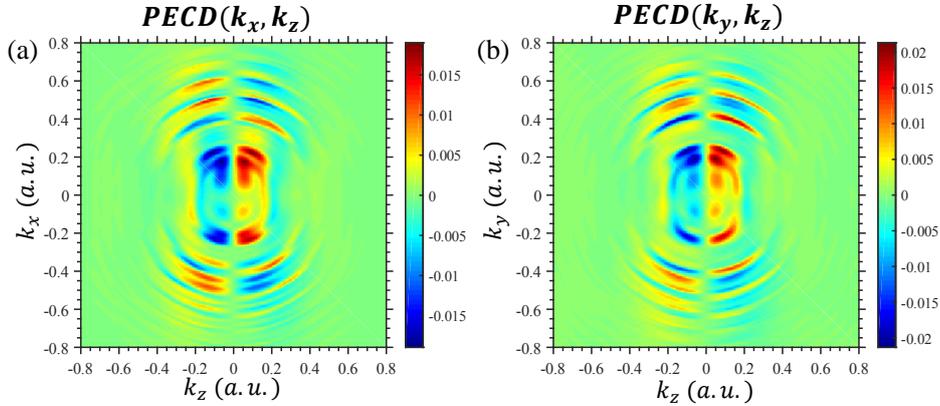

FIG. 6. PECD using CPL from CBrClFH. Calculations are performed by substituting $\alpha=0$, $\Delta=0$, $\varepsilon_1=1$ in eq. (1), and setting the laser power to $I_0=4\times10^{13}$W/cm$^2$, such that the total power is equal to that in Fig. 2 in the main text that uses ω-2ω. Note that the color scales are different in all sub-plots.

Next, we present ATI CD calculations obtained from a higher laser power of $I_0=4\times10^{13}$W/cm$^2$, as compared to results in Fig. 3 in the main text. Fig. 7 presents the ATI CD and generally shows a stronger CD with a maximal value of 7.04%. Note that even though the maximal value of the ATI CD is larger, the total ionization rate CD here is 0.31%, which is lower than its value of 2.46% at the lower laser power because the ATI CD is more oscillatory and changes sign between different ATI peaks. This is an indication that it is not necessarily straightforward that higher laser powers directly lead to stronger chiral signals. We note that by selectively removing photoelectrons with low energies (or high energies) one can artificially increase the total ionization rates CD.

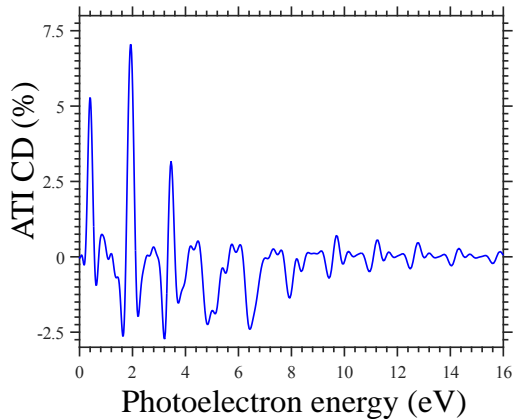

FIG. 7. ATI CD from CBrClFH at a higher laser power. Calculations are performed in similar ω-2ω optical beam settings to those in Fig. 3, but with a larger laser power of $I_0=4\times10^{13}$W/cm$^2$.

# APPENDIX C: ADDITIONAL RESULTS FROM LIMONENE, FENCHONE, AND CAMPHOR

We present here additional results from the chiral molecules Limonene, Fenchone, and Camphor. Fig. 8 presents the PECD in all hemispheres calculated for these molecules in the same settings as Fig. 5 in the main text. Limonene here shows the strongest CD, in accordance with the ATI CD presented in Fig. 5.

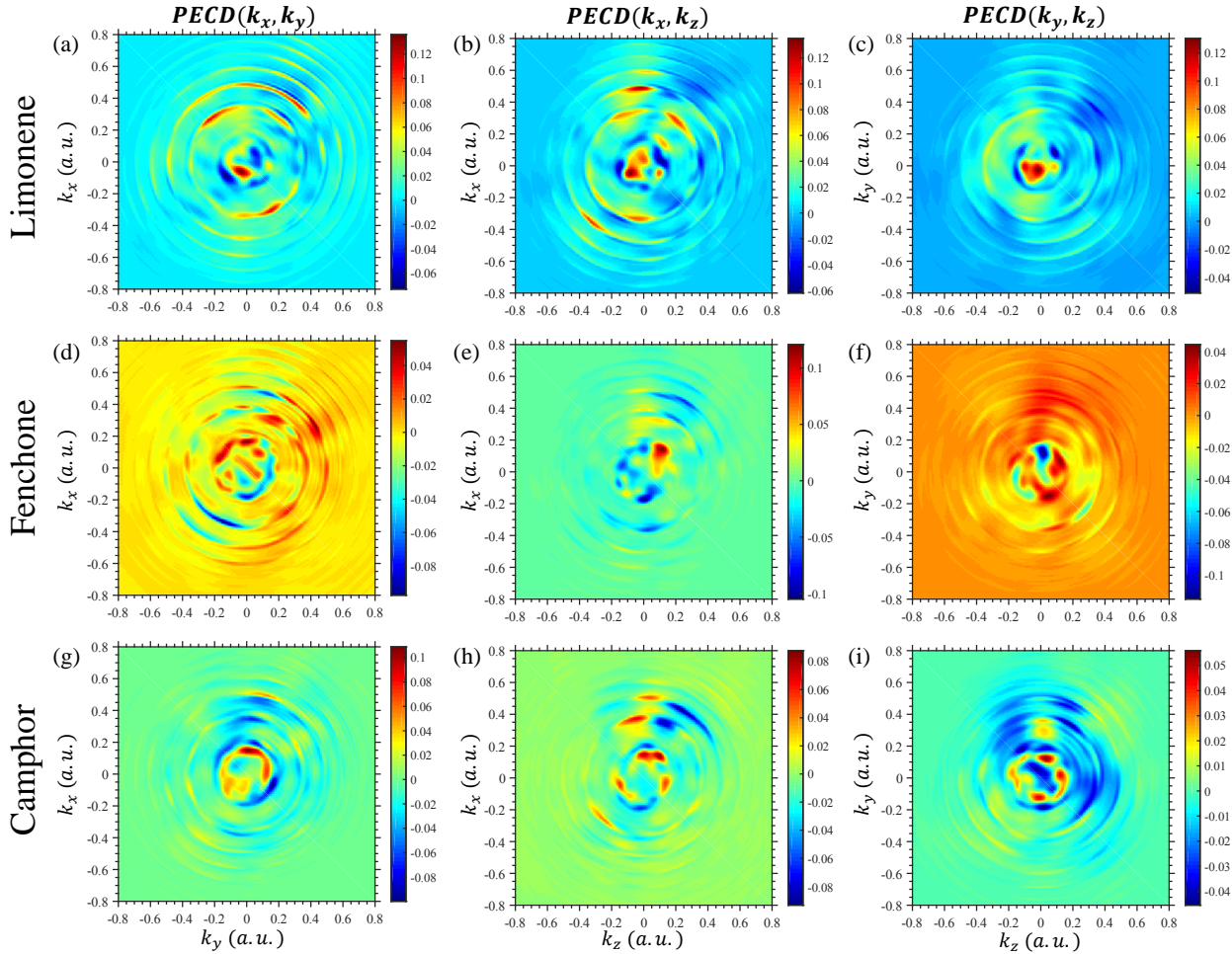

FIG. 8. PECD spectra from (a-c) Limonene, (d-f) Fenchone, (g-i) Camphor, respectively. Calculations are performed in similar ω-2ω optical beam settings to those in Fig. 5 in the main text. Note that the color scales are different in all sub-plots.

---